\documentstyle[12pt,a4wide]{article}

\input{epsf}

\def\IR{\relax{\rm I\kern-.18em R}}
\def\IN{\relax{\rm I\kern-.18em N}}

\def\r{\rho}
\def\H{{\cal L}}
\def\O{{\cal O}}

\newcommand{\be}{\begin{equation}} \newcommand{\ee}{\end{equation}}
\newcommand{\bea}{\begin{eqnarray}} \newcommand{\eea}{\end{eqnarray}}

\begin{document}

\title{Angular quantization and the density matrix
  renormalization group}

\author{J. Gaite%
\footnote{Also at {\it Instituto de Matem{\'a}ticas y F{\'\i}sica
Fundamental, CSIC, Serrano 123, 28006 Madrid, Spain}}\\ {\it Centro de
Astrobiolog{\'\i}a\footnote{Associated to NASA Astrobiology Institute}, 
INTA-CSIC,}\\
{\it Ctra.\ de Torrej\'on a Ajalvir,}
{\it 28850 Torrej\'on de Ardoz, Madrid, Spain.}
}

\maketitle

\begin{abstract}
Path integral techniques for the density matrix of a
one-dimensional statistical system near a boundary previously 
employed in black-hole physics are applied to providing a 
new interpretation of the density matrix
renormalization group: its efficacy is due to the concentration of 
quantum states near the boundary. 
\end{abstract}


There has recently been a revival of interest in the study of 
statistical properties of
quantum field theories in the presence of a boundary.
This subject has a long history: it goes back (at least) to the idea
of the entropy of black holes. This entropy, associated to Hawking radiation, 
was studied with
quantum field
theory in the corresponding curved space-time \cite{Un}. 
It has been proposed that black-hole entropy arises solely
from the presence of a horizon and the consequent ignorance of
its interior, being an {\em entanglement} entropy and having
little to do with the curvature of the space-time \cite{Bomb}. Actually, 
this is indicated by the classical calculation of the density matrix 
in Rindler space \cite{Un}, which will be important in the following. 

An interesting connection arose from the study of integrable 
one-dimensional lattice models, where R.~Baxter introduced the {\em corner 
transfer matrix}. In the continuum limit of these models, it is natural 
to use rapidity variables, in which the integrability conditions 
adopt a simpler form and, moreover, 
their symmetry with respect to phase shifts 
can be understood as a consequence of Lorentz invariance \cite{Thac}. 
Substantiating this connection,  
the type of quantization used in the Euclidean version of
Rindler space, namely, {\em angular quantization}, has been
invoked as a suitable computational formalism in 1+1 integrable
quantum field theory \cite{BraLuk}.

The corner transfer matrix method is only appropriate for 
integrable models, but T.~Nishino and collaborators realized that it 
is related to an approximate method to solve quantum systems, namely, 
the {\em density matrix renormalization group} (DMRG) \cite{Nishi}.

The DMRG is a numerical renormalization group introduced by S. White
\cite{White}, which seems to provide unprecedented precision. 
White intended to tackle the problem of the influence of
boundary conditions in the application of the {\em real space} 
RG to finite systems.
A first attempt on accounting for the effect of boundary conditions was
the combination of boundary conditions approach \cite{WhiNo}.
It eventually led to the DMRG, by appealing to
Feynman's formulation of the density matrix as the {\em best}
description of a part of a quantum mechanical system: 
the DMRG method then reduces the number of
states of the subsystem by discarding the smallest eigenvalues of the
density matrix.
The DMRG is analogous to Wilson's treatment of the Kondo
problem in that, at every step, new states are added at one side of
the system and then other states are removed to keep the size of the
Hilbert space approximately constant.  It seems that, in this process,
it is sensible to keep more states near the boundary. 
This is reminiscent of the concentration of quantum states at 
a black-hole's horizon.  
We shall see that the angular quantization construction of the density
matrix shows that the DMRG precisely amounts to an algorithm to keep
more states near the boundary in a systematic way.

Before proceeding, it is interesting to mention that there is a
well-known condensed matter system whose dynamics can be 
considered as concentrated on a spatial boundary, namely, the 
two-dimensional electron system exhibiting the quantum Hall effect. 
Its connection with the physics of black holes has already 
been worked out \cite{Bala}.

The type of models to which the DMRG is usually applied consists of
those defined on a chain, such as the 1D Hubbard or Heisenberg model,
the Ising model in a transverse field, etc.  For our purposes, it is
convenient to consider simpler models and we shall illustrate the
effect of the DMRG on a chain of oscillators.  Of course, if it is a
harmonic chain, its Hamiltonian can be diagonalized in momentum space,
yielding a spectrum of free phonons, and no renormalization group is
necessary. Nevertheless, it constitutes a suitable test for the DMRG
\cite{CaMou}, since this RG acts in real space, where the kinetic 
term is a coupling in its own right. 

The density matrix of {\em half} of the harmonic chain is a straightforward 
generalization of the density matrix of a harmonic oscillator coupled to the 
environment and can be
obtained by the diagonalization of its Hamiltonian, the construction of the 
ground state, and a subsequent
trace over the unobservable variables
\cite{Bomb,Pes}. However, it is preferable
to use path integral methods in the continuum, which are simpler and
allow one to obtain more general results \cite{KabStr}.

Therefore, we consider a chain of coupled
oscillators, which gives rise 
to phonon-like collective excitations.
In the continuum limit, the action for this model is 
\be
S[u(x,t)] = \int\! dt\int\! dx \left({\mu\over 2}\left[(\partial_tu)^2 - 
c^2\,(\partial_x u)^2 \right] - V(u) \right),
\ee
where $u$ is the displacement, $\mu$ is the mass per unit length, and 
$c$ is the speed of sound. 
The Hamiltonian can be expressed as the integral of the energy density, 
which is the time-component of the energy-momentum tensor:
\be
H = \int T_{00}\,dx.
\label{H}
\ee
After redefining $t$ and $u$ to have $xt$-symmetry and to remove $\mu$,
\be
T_{00} = {1\over 2} \left[(\partial_tu)^2 + (\partial_x u)^2 \right] + 
V(u).
\label{too}
\ee


Let us obtain a path integral representation for the density matrix 
on the half-line of a system that is in its ground state \cite{KabStr}.  
We use the notation corresponding to the oscillator chain but it shall 
be obvious that the construction is fairly general. 
In the continuum limit, the half-line density matrix is a functional integral,
\be
\r[u_R(x),u'_R(x)] = \int\! Du_L(x)\, \psi_0[u_L(x),u_R(x)]\,
\psi^*_0[u_L(x),u'_R(x)],
\label{DM0}
\ee
where the subscripts refer to the left or right position of the
coordinates with respect to the boundary (the origin). Now, we must
express the ground-state wave-functions as a path integral, 
\be
\psi_0[u_L(x),u_R(x)] = \int Du(x,t)\, \exp\left(-S[u(x,t)]\right),
\ee
where $t\in (-\infty,0]$ and with boundary conditions $u(x,0) =
u_L(x)$ if $x<0$, and $u(x,0) = u_R(x)$ if $x>0$.  The conjugate
wave function is given by the same path integral and boundary
conditions but with $t\in [0,\infty)$.  Substituting into
Eq.~(\ref{DM0}) and performing the integral over $u_L(x)$, one can
express $\r(u_R,u'_R)$ as a path integral over $u(x,t)$, with $t\in
(-\infty,\infty)$, and boundary conditions $u_R(x,0+) = u'_R(x)$,
$u_R(x,0-) = u_R(x)$.  In other words, $\r(u_R,u'_R)$ is represented
by a single path integral covering the entire plane with a cut along
the positive semiaxis, where the boundary conditions are imposed.

Let us now consider the {\em angular} evolution operator $\exp
(-2\pi\H)$, where $\H$ is the generator of rotations around the origin
and $2\pi$ is the angle's range, in analogy with the canonical density
matrix $\exp (-\beta\,H)$.  Its matrix element in the Schr\"odinger
representation $\langle u'_R |\exp (-2\pi\,\H) |u_R\rangle$ is given
by an angular path integral with boundary conditions $u_R(0) = u'_R$
and $u_R(2\pi) = u_R$, in analogy with the canonical density matrix
path integral. Therefore, it precisely coincides with the density
matrix path integral.

The DMRG algorithm discards the smallest eigenalues of the density
matrix and is, therefore, equivalent to truncating the spectrum of $\H$. 
Of course, this truncation has the typical variational flavor of real 
space renormalization groups, and we can say that it selects  
states adequate to the presence of the boundary.   

In Euclidean two-dimensional field theory, 
the generator of rotations in the $(x,t)$ plane is 
\be
\H =  \int dx\,(t\,T_{11}-x\,T_{00}).
\label{L}
\ee
To simplify, one can evaluate it at $t=0$. For the oscillator chain, 
we must use the value of $T_{00}$ (\ref{too}). In the Schr\"odinger
representation, we should replace the momentum $\Pi = \partial_tu$ 
with $\Pi(x) = i\,\delta/\delta u(x)$.  However, as in canonical quantization,
one rather uses the second-quantization method, which diagonalizes the
Hamiltonian by solving the classical equations of motion and
quantizing the corresponding normal modes.  
Let us recall that, in canonical quantization, 
if we disregard anharmonic terms, 
the classical equations of motion in the continuum limit become 
the Klein-Gordon field equation, giving rise to the usual
Fock space.  Not surprisingly, the eigenvalue equation for $\H$ leads
to the Klein-Gordon equation in polar coordinates in the $(x,t)$
plane,
\be
(\Delta + m^2) u = 
\left({1\over r}{\partial \over \partial r}r{\partial \over \partial r} +
{1\over r^2}{\partial^2 \over \partial \phi^2}+ m^2\right) u = 0.
\ee
Separating the angular variable, it becomes a Bessel differential
equation in the $r$ coordinate with complex solutions $I_{\pm
i\,\ell}(m\,r)$ \cite{GrRy}, $\ell$ being the angular frequency. We
have a continuous spectrum, which becomes discrete on introducing
boundary conditions. One of them must be set at a short distance from
the origin, to act as an ultraviolet regulator \cite{Bomb,KabStr}, 
necessary in the continuum limit.

Therefore, the second-quantized field is (on the positive semiaxis 
$t=0 \Leftrightarrow \phi = 0$, $x \equiv r$)
\be
u(x) = \int {d\ell\over 2\pi}\,\frac{b_\ell\,I_{i\,\ell}(m\,x) + 
b_\ell^\dag\,I_{-i\,\ell}(m\,x)}{\sqrt{2\,\sinh(\pi\,\ell)}},
\ee
where we have introduced annihilation and creation operators and 
where the term that appears in the denominator is just for normalization, 
to ensure that those operators satisfy 
canonical conmutations relations.
There is an associated Fock space built by acting with $b_\ell^\dag$
on the ``vacuum state''. These states constitute the spectrum of
eigenstates of $\H$, which adopts the form 
$\H = \int d\ell \,\ell\, b_\ell^\dag b_\ell$ (where the integral is 
replaced with a sum for discrete $\ell$).

We proceed to show, by using the eigenfunctions of $\H$ instead of
free waves, that we have a basis in which the region close to $x=0$---the
boundary point---is more accurately represented than the region far
from it when we cut off the higher $\ell$ eigenfunctions.  Notice that the
functions $I_{\pm i\,\ell}(m\,x)$ have wave-lengths that increase with
$x$.  It is illustrative to represent a real solution,
$$K_{i\,\ell}(m\,x) = \frac{i\,\pi}{2\,\sinh(\pi\,\ell)}\,
[I_{i\,\ell}(m\,x)-I_{-i\,\ell}(m\,x)].$$
A detailed analysis shows that this solution is oscillatory for $x < \ell/m$, 
with a wavelength proportional to $x$, and decays 
exponentially for $x > \ell/m$ (Fig.~1). 
The other type of real solution, $I_{-i\,\ell}(m\,x)+I_{i\,\ell}(m\,x)$, 
has the same behavior for $x < \ell/m$ but for $x > \ell/m$ grows
exponentially instead.
For unattenuated phonons ($m=0$), there is 
no exponential decay or growth. 
Imposing a maximum value of $\ell$, we restrict the amount of detail
that a linear combination of these functions can have but, clearly,
the allowed amount of detail is larger for the region close to the origin than
for the region far from it.
In the second-quantized theory, the $\ell$ cutoff has a similar 
interpretation: 
near the origin, a larger range of energy of modes (phonons) is allowed,
so representing the influence of the other half of the chain.

\begin{figure}
  \epsfxsize=8cm 
\epsfbox{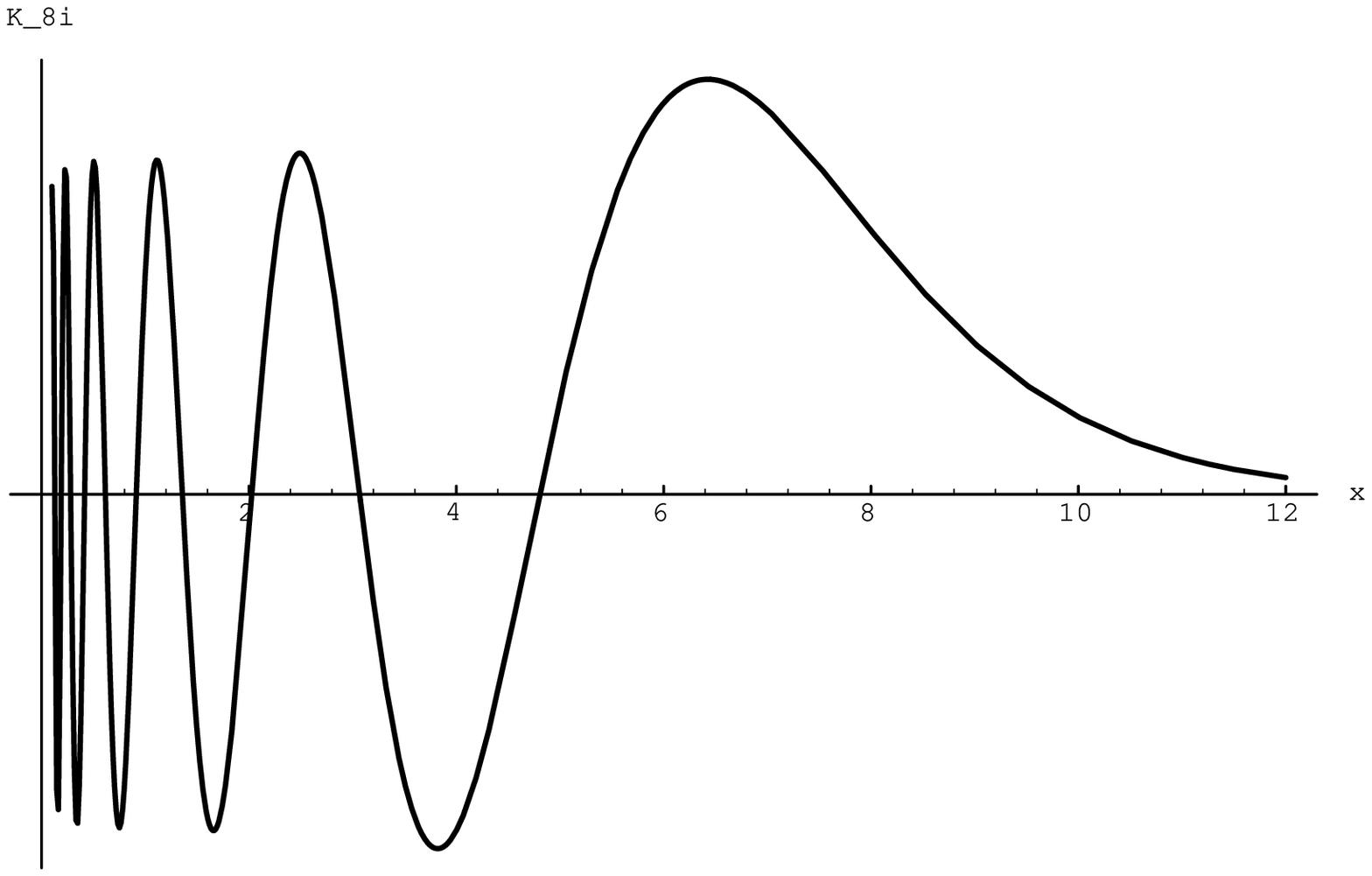}
\flushleft
{Fig.~1. The real solution $K_{8\,i}(x)$. Its exponential decay is apparent for $x > 8$.}
\vskip .5cm 
\end{figure}

Let us see how the numerical DMRG would actually proceed for the chain
of coupled harmonic oscillators (see also Ref.~\cite{CaMou}). 
Since the Hilbert space for a single
oscillator already has infinite dimension, one must begin by
truncating the initial Hilbert space to a finite and rather small
number of states {\em per site}, $n$ say.  
Then one obtains the ground state, with energy
$(1/2)\sum_{-N}^{N}\omega_k$, $\omega_k$ being the frequencies of the
$N$ normal modes $a_k$ (of ordinary canonical quantization). 
From the ground state, one derives the corresponding density
matrix of half of the chain. This can be done numerically, but
analytically as well, since the integral for the density matrix is
Gaussian [see Eq.~(\ref{DM0})].
In White's original algorithm \cite{White}, one begins
with only one site, which is reflected on the origin, to compute the
two-site ground state and the one-site density matrix. 
Next, one discards its smallest eigenvalues and iterates: one adds one
site at the center with their $n$ new states, reflects, etc.  In the
present case, the one-site density matrix is Gaussian and therefore 
equivalent to the thermal matrix of one single oscillator, 
with a temperature 
increasing with the coupling strength. Thus it has the form 
$\exp(-2\pi\H)$, where $2\pi\H$ is $\beta\,H$ for 
the ``equivalent thermal oscillator'' \cite{Pes}. 

For even $N\geq 2$, the integral (\ref{DM0}) is an ordinary integral in $N/2$
variables. It yields the exponential of a quadratic form, $u_R^{T} A
u_R+{u'_R}^{T} A u'_R + {u_R}^{T} B u'_R + {u'_R}^{T} B u_R$, where
$A$ and $B$ are $N/2 \times N/2$ symmetric matrices, the former being
positive definite.  One can then perform a two-step diagonalization of
this quadratic form \cite{Gold}: If we write $A = M\,M^T$, with $M$ a
$N/2 \times N/2$ nonsingular real matrix 
(which is always possible \cite{GrRy}), the
transformation $v = M^Tu$ puts the first part in canonical form, while
changing $B \rightarrow M^{-1}B\left({M^{-1}}\right)^T$. This matrix can be
diagonalized by an orthogonal transformation $\O$, without altering
the canonical form of the other part. The result is that the density
matrix becomes the product of density matrices of independent modes.
Consequently, it adopts the form $\exp (-2\pi\H)$, with $2\pi\H$
being the sum of the operators $\beta\,H$ for those modes
(the equivalent thermal oscillators).
Furthermore, the total transformation matrix $\O^T M^T$ is the discrete
version of the functions $K_{i\,\ell}(m\,x)$, such that the first
index corresponds to $\ell$ while the second corresponds to $x$.  We
have plotted in Fig.~2
the second and third file vectors of that matrix, for the case $m=0.1$ and 
$N=24$
.  It can be seen that they are a discrete version of functions of the
type of that in Fig.~1, although with fewer oscillations. The
following file vectors correspond to higher $\ell$, and hence to more
rapidly oscillating functions, but the discreteness interferes in such
a manner that these oscillations are difficult to perceive.

\begin{figure}
  \epsfxsize=8cm \epsfbox{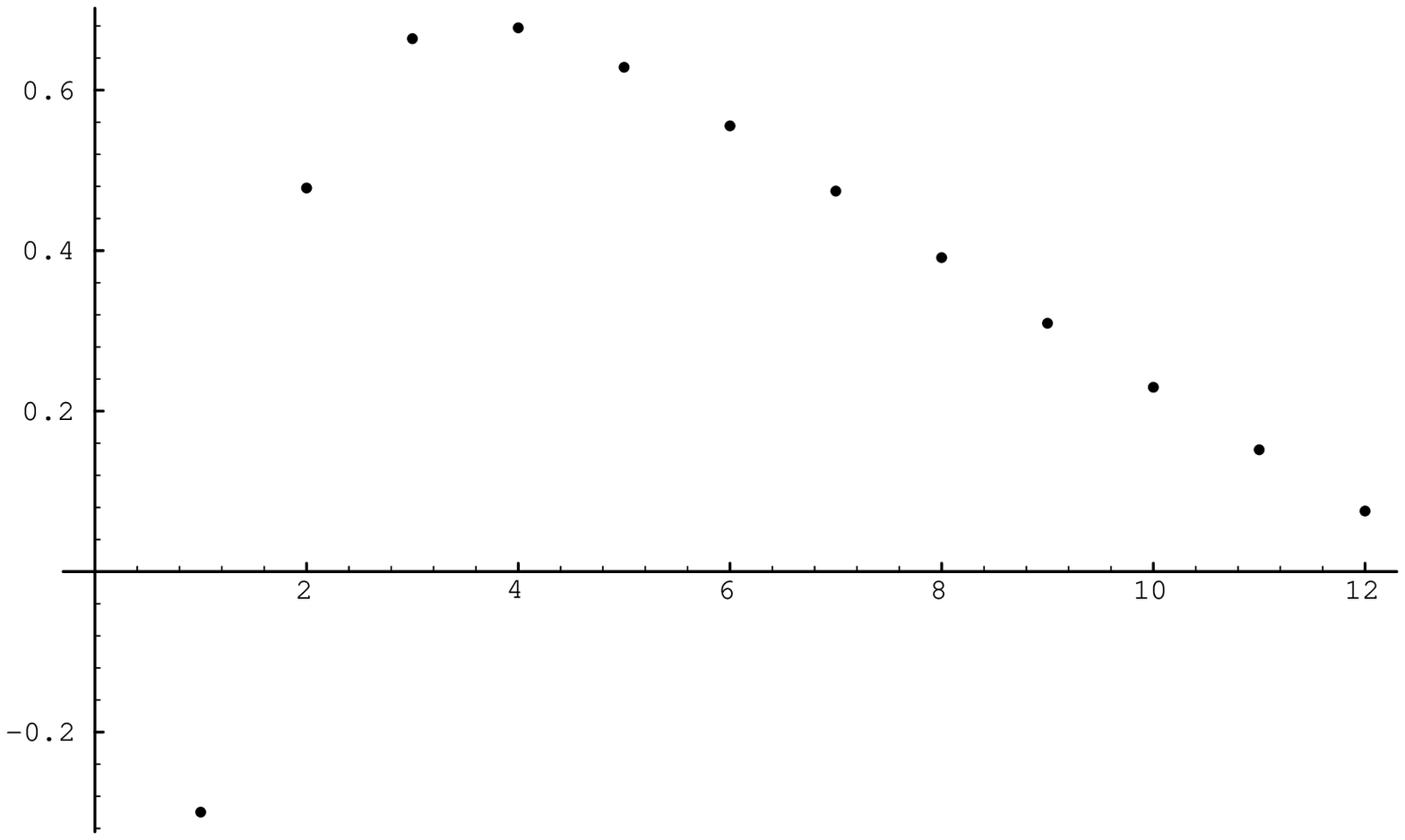} 
  \epsfxsize=8cm \epsfbox{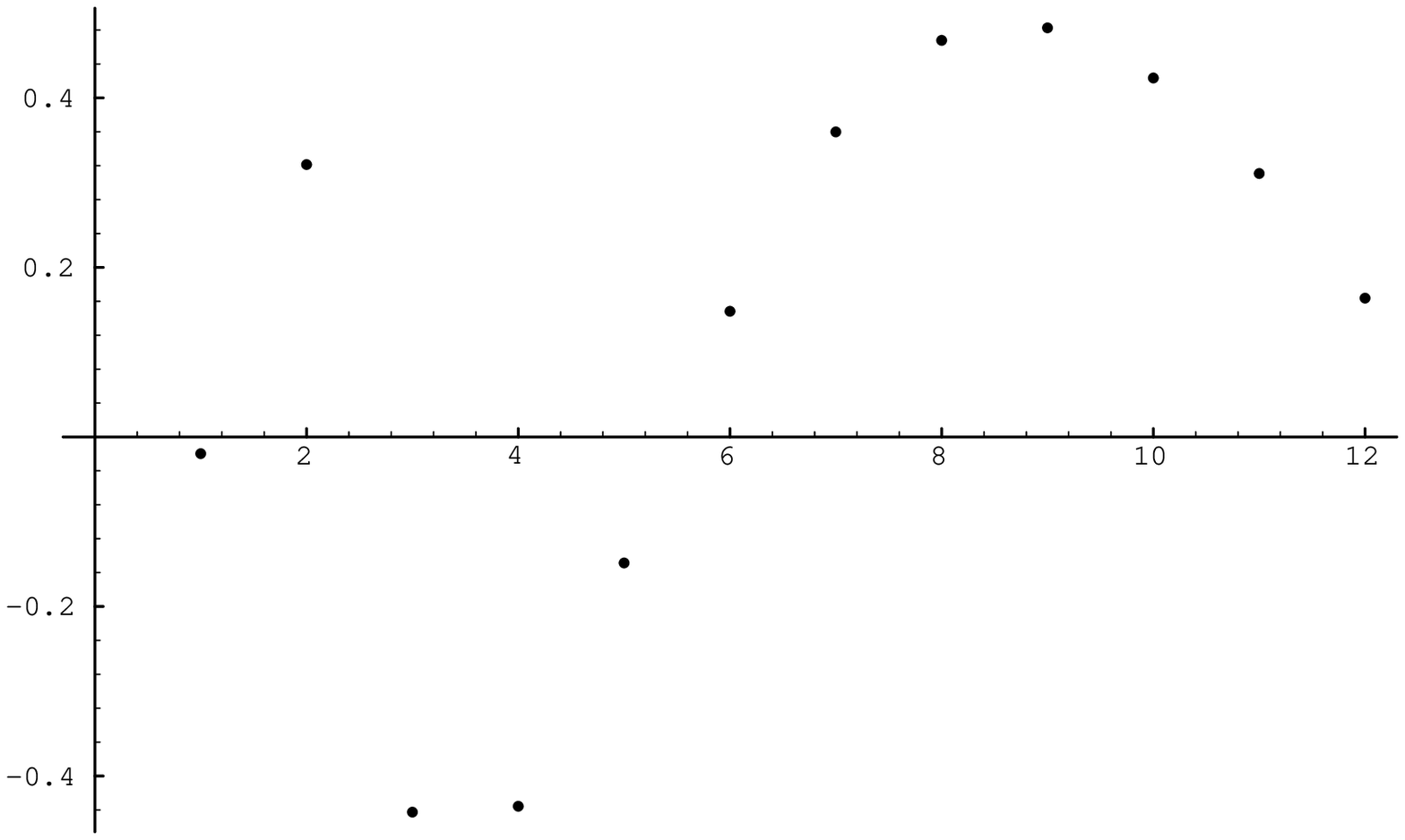}
{Fig.~2. Plots of the second and third line vectors of
  the matrix of transformation to normal mode variables.}  
\end{figure}

The inclusion of weak anharmonic terms in the action should not
substantially modify the picture above. 
As mentioned before, in real space the important coupling is actually
the kinetic term, which couples 
neighboring points. As regards to calculation, anharmonic terms 
prevent us from obtaining analytic expressions for the angular states 
or performing the functional integrals, which are no longer Gaussian. 
However, the effect of those terms amounts to a
mode-mode coupling, which can be treated perturbatively. 
Canonical methods for perturbation theory in Rindler space 
are exposed in Ref.~\cite{Un}. On the other hand, it is easy to 
see how to evaluate the functional integral for the density matrix, 
Eq.~(\ref{DM0}), in perturbation theory: One must first introduce 
the coupling to an external current, as usual, and express 
the interaction term, $V(u) = \lambda\,u^4$ say, as a derivative with respect 
to the external current; then, the Gaussian integral can be performed
and the perturbation series constructed. 
At any rate, although these perturbative corrections 
modify the ground and excited-state energies, the
qualitative properties of the spectrum of $\H$ must remain the same.

We must remark that, in
quantum field theory in $1 + 1$ dimensions, {\em
non-perturbative} phenomena may occur and one must
have an idea about the type of spectrum of collective excitations
before applying a perturbative approach, since these effects can turn a 
bosonic spectrum into a fermionic one (or viceversa) \cite{Abda}.
If the collective
excitations are bosons, the formulas above would constitute the basis
for perturbation theory and the conclusions should not be modified.
On the other hand, it should not be difficult to reach analogous
conclusions for a fermionic spectrum such as, for example, the
spectrum of the Ising model in a transverse field, which has been the
subject of many tests for the DMRG. The field-theory forms of $H$ and
$\H$ given by Eq.~(\ref{H}) and (\ref{L}), respectively, still
hold. Their particular expressions for the free fermion field are easy
to obtain.  The form of the angular wavefunctions would be similar to 
the bosonic ones, already studied.

Moreover, regarding the generality of our conclusions, one can argue
that the essence of the DMRG, as concerns the effect of boundary
conditions, is independent of the type of spectrum or the details of
perturbation theory.  In a general 1D many-body system that is
homogeneous in the continuum limit, such as a system defined on a
chain with uniform site-to-site coupling, we can always express the
energy by the integral (\ref{H}) and, therefore, the generator of
rotations as $\H = \int dx\,x\,\epsilon(x),$ where $\epsilon :=
T_{00}$ does not depend explicitly on $x$.  Suppose that we divide the
chain into equal-size blocks, with coordinate $x_j$, such that, for
sufficiently large blocks, we can approximately write $\H = \sum_j
x_j\,\epsilon_j$, except for a small inter-block potential.  If we
want to discard eigenvalues of the density matrix $\exp(-2\pi\H)$
smaller than certain value, then we are to set an upper cutoff for the
$\H$ eigenvalues, which then implies a cutoff for every $\epsilon_j$,
but depending inversely on $j$: for small $j$, that is, close to the
boundary, an extensive range of values is included, accounting for the
progressive uncertainty in their value as the boundary is approached,
whereas for large $j$ only the lowest energy states are allowed.  In
conclusion, one necessarily has a concentration of quantum states at
the boundary, due merely to the existence of this boundary, as in the
different physics of black holes.

Finally, let us mention that, for a class of integrable models, one can go
beyond perturbation theory and, actually, calculate the correlation
functions from the two-body $S$-matrix.  Owing to its relation with 
the corner transfer matrix, the
{\em form-factor} method to calculate the correlation functions for
these models is deeply related with angular quantization
\cite{BraLuk}.  In this context, it would be very interesting to
compare the results of the DMRG for these models with those of the
form-factor approach.



\begin{thebibliography}{99}



\bibitem{Un} N.D. Birrell, P.C.W. Davies, {\em Quantum fields in 
curved space}, Cambridge U.P. (1982)

\bibitem{Bomb} L. Bombelli, R.K. Koul, J. Lee and R.D. Sorkin,
Phys.~Rev.~{\bf D~34} (1986) 373--383; M. Srednicki,
Phys. Rev. Lett. {\bf 71} (1993) 666

\bibitem{Thac} H.B.~Thacker, Physica~{\bf D~18} (1986) 348

\bibitem{BraLuk} V. Brazhnikov, S. Lukyanov, Nucl.~Phys.~{\bf B~512}
(1998) 616--636


\bibitem{Nishi} T. Nishino and K. Okunishi, {\em Corner Transfer
Matrix Algorithm for Classical Renormalization Group}, {\tt
cond-mat/9705072}, and references therein

\bibitem{White} S.R. White, Phys. Rev. Lett. {\bf 69} (1992) 2863; Phys.
  Rev. {\bf B 48} (1993) 10345

\bibitem{WhiNo} S.R. White and R.M. Noack, Phys. Rev. Lett. {\bf 68}
(1992) 3487

\bibitem{Bala} A.P. Balachandran, L. Chandar, A. Momen, talk given at
the 17th Annual Montreal-Rochester-Syracuse-Toronto Meeting on High
Energy Physics, Rochester, NY, 8--9 May 1995, {\tt gr-qc/9506006};
Nucl.\ Phys.~{\bf B 461} (1996) 581--596; Int.\ J.\ Mod.\ Phys.~{\bf A
12} (1997) 625--642

\bibitem{CaMou} L.G. Caron and S. Moukouri, Phys. Rev. {\bf B 56} (1997) 8471

\bibitem{Pes} D. Han, Y.S. Kim and M.E. Noz, Am. J.~Phys. {\bf 67}
(1999) 61; I. Peschel, M.-C. Chung, J.~Phys.~A: Math.~Gen. {\bf 32}
(1999) 8419--8428

  
  
\bibitem{KabStr} D. Kabat and M.J. Strassler, Phys. Lett. {\bf B 329}
(1994) 46--52; C. Callan and F. Wilczek, Phys. Lett. {\bf B 333}
(1994) 55--61

\bibitem{GrRy} I.S. Gradshteyn and I.M. Ryzhik, {\em Table of
    integrals, series, and products}, Academic Press, (1980)

\bibitem{Gold} This procedure is analogous to the diagonalization of
  the Hamiltonian in the theory of small oscillations; see H.
  Goldstein, {\it Classical Mechanics}, Second Edition,
  (Addison-Wesley, 1984). It has been used in Ref.~\cite{Bomb}.

  
\bibitem{Abda} There are several treatises on soluble models and
non-perturbative phenomena in $1 + 1$ dimensions. The point of view of
many-body theory is exposed in, for example, D.C. Mattis, {\em The
many body problem}, World Scientific, Singapore, (1993), while the
field-theory point of view is exposed in, for example, E. Abdalla,
M.C.B. Abdalla and K.D. Rothe, {\em 2~Dimensional Quantum Field
Theory}, World Scientific, Singapore, (1991).
  
\end{thebibliography}
\end{document}